\documentclass[twocolumn,usenatbib,usegraphicx ]{mn2e}

\def\Mpc{{\rm Mpc}}
\def\lcdm{$\Lambda {\rm CDM} \,\,$} 
\def\mnras{MNRAS}

\def\aj{AJ}
\def\apj{ApJ}
\def\apjl{ApJL}
\def\apjs{ApJS}
\def\aap{A\&A}

\usepackage{epsfig}

\begin{document}
\title{Statistically significant length scale of filaments as a robust
  measure of galaxy distribution} \author[B. Pandey ] {Biswajit
  Pandey$^1$\thanks{Email: biswap@visva-bharati.ac.in}
  $^2$\thanks{Email: biswa@iucaa.ernet.in} \\ ${}^1$ Department of
  Physics, Visva-Bharati, Santiniketan, Birbhum, 731235, India
  \\ ${}^2$ Inter-University Centre for Astronomy and Astrophysics,
  Post Bag 4, Ganeshkhind, Pune, 411 007, India \\ } \maketitle

\begin{abstract}

We have used a statistical technique ``Shuffle" \citep {bhav, bharad2}
in seven nearly two dimensional strips from the Sloan Digital Sky
Survey Data Release Six (SDSS DR6) to test if the statistically
significant length scale of filaments depends on luminosity, colour and
morphology of galaxies. We find that although the average
filamentarity depends on these galaxy properties, the statistically
significant length scale of filaments does not depend on them. We
compare it's measured values in SDSS against the predictions of
$\Lambda$CDM N-body simulations and find that $\Lambda$CDM model is
consistent with observations. The average filamentarity is known to be
very sensitive to the bias parameter. Using $\Lambda$CDM N-body
simulations we simulate mock galaxy distributions for SDSS NGP
equatorial strip for different biases and test if the statistically
significant length scale of filaments depends on bias. We find that
statistically significant length scale of filaments is nearly
independent of bias. This result is possibly related to the fact that
statistically significant length scale of filaments is nearly the same
for different class of galaxies which are differently biased with
respect to underlying dark matter distribution. The average
filamentarity is also known to be dependent on the galaxy number
density and size of the samples. We use $\Lambda$CDM dark matter
N-body simulations to test if the statistically significant length
scale of filaments depends on number density of galaxies and size of
the samples. Our analysis shows that the statistically significant
length scale of filaments very weakly depends on these
factors. Finally we test the reliability of our method by applying it
to controlled samples of segment Cox process and find that our method
successfully recovers the length of the inputted segments. Summarizing
these results we conclude that the statistically significant length
scale of filaments is a robust measure of the galaxy distribution.

\end{abstract}

\section{Introduction}

The fact that the galaxies appear to be distributed along filaments
which are interconnected to form a web like structure which is often
reffered as the `cosmic web' is one of the most striking visual
feature in all the present and past redshift surveys (e.g. , CfA ,
\citealt{gel}; LCRS, \citealt{shect}; 2dFGRS, \citealt{colles} and
SDSS, \citealt{stout}).

The analysis of filamentary patterns in the galaxy distribution has a
long history dating back to a few papers in the late-seventies and
mid-eighties by \citet{joe}, \citet{einas4}, \citet{zel},
\citet{shand1} and \citet{einas1}. Filaments are the most striking
visible patterns seen in the galaxy distribution (e.g. \citealt{gel},
\citealt{shect}, \citealt{shand2}, \citealt{bharad1}, \citealt{mul},
\citealt{basil}, \citealt{doro2}, \citealt{pimb}, \citealt{pimb1},
\citealt{pandey}). A review on a number of physically motivated and
statistical methods to define filaments is provided in \citet{pimb2}.
The percolation analysis (eg. \citealt{shand1}, \citealt{einas1}), the
genus statistics (eg. \citealt{gott}), the minimal spanning tree
(e.g. \citealt{barrow}), the Voronoi tessellation (\citealt{ike},
\citealt{weygaert}), the Minkowski functionals (eg. \citealt{mecke},
\citealt{smal}) and the `Shapefinders' \citep{sahni} are some of the
useful statistical tools introduced to quantify the topology and
geometry of the galaxy distribution. \citet{stoi1} propose to apply a
marked point process to automatically delineate filaments in the
galaxy distribution. \citet{colberg} studied the intercluster
filamentary network in high resolution N-body simulations of structure
formation in a $\Lambda$CDM Universe. \citet{arag} use Multiscale
Morphology Filter technique to identify wall-like and filament-like
structures in cosmological N-body simulations.  \citet{sus} propose a
skeleton formalism to quantify the filamentary structure in three
dimensional density fields. \citet{stoi2} propose to apply an object
point process to objectively identify filaments in galaxy redshift
surveys. \citet{sarkar1} propose the Local Dimension to locally
quantify the shape of large scale structures in the neighbourhood of
different galaxies in the Cosmic Web.

The SDSS \citep{york} is currently the largest galaxy redshift
survey. In an earlier work \citep{pandey} (hereafter Paper I) we have
analysed the filamentarity in the equatorial strips of this
survey. These strips are nearly two dimensional and we have projected
the data onto a plane and analysed the resulting 2-D galaxy
distribution. We find evidence for connectivity and filamentarity in
excess of that of a random point distribution, indicating the
existence of an interconnected network of filaments. We find that
filaments are statistically significant upto length scales $80 \,
h^{-1} {\rm Mpc}$ and not beyond \citep{pandey}. All the structures
spanning length-scales larger than this length scale are the result of
chance alignments. This is consistent with an earlier analysis by
\citet{bharad2} where they show that in Las Campanas Redshift Survey
(LCRS) the largest length-scale at which filaments are statistically
significant is between $70$ to $80 \, h^{-1}$Mpc. Further we show that
the average filamentarity of the galaxy distribution depends on various
physical properties of galaxies such as luminosity, colour, morphology
and star formation rate \citep{pandey1, pandey3}. It would be
interesting to know if the statistically significant length scale of
filaments also depends on different galaxy properties. In the present
work we study if the statistically significant length scale of
filaments depends on luminosity, colour and morphology of galaxies.

Further it is also possible to measure the statistically significant
length scale of filaments in mock galaxy distribution extracted from
N-body simulation.  The \lcdm model is currently believed to be the
minimal model which is consistent with most cosmological data
(\citealt{efst}; \citealt{perci}; \citealt{teg1}; \citealt{sperg};
\citealt{sperg1}; \citealt{komat}). It would be interesting to measure
the statistically significant length scale of filaments in N-body
simulations of $\Lambda$CDM model and compare it with measured values
in the SDSS \citep{pandey}. The N-body simulations primarily predict
the clustering of the dark matter. This should be contrasted with the
fact surveys reveal only the bright side of the matter
distribution. Galaxy formation is a complicated process and the exact
relation between the distribution of the galaxies and the dark matter
is not well understood. It is now generally accepted that the galaxies
are a biased tracer of the dark matter distribution (e.g.,
\citealt{kais}; \citealt{mo}) and on large scales one expects the
fluctuations in the galaxy and the dark matter distribution to be
linearly related through the linear bias parameter $b$. Mock galaxy
distributions with different bias can be simulated following this
assumption. The average filamentarity of the simulated galaxy
distribution is found to be very sensitive to the bias parameter
\citep{bharad3, pandey2}. It would be interesting to know how the
statistically significant length scale of filaments depends on the
bias parameter. In the present work we test if the statistically
significant length scale of filaments depends on bias.

Earlier we find that the average filamentarity is sensitive to the
area and galaxy number density of the samples
\citep{pandey1}(hereafter Paper II) and this statistics can be used
for a meaningful comparison between two different galaxy samples only
when they have the same volume (identical shape and size) and galaxy
number density. In the present work we would also like to test if the
statistically significant length scale of filaments depends on area
and galaxy number density of the samples.

Finally we simulate segment Cox process \citep{pons} with different
segment lengths and test the efficiency of our method in measuring the
statistically significant length scale of filaments in the mock
samples drawn from these simulations.

A brief outline of our paper follows. Section 2 describes the data and
method of analysis, our results and conclusions are presented in
Section 3.

\section{Data and Method of analysis}


\subsection{SDSS data}

The Sloan Digital Sky Survey (SDSS) \citep{york} is a wide-field
imaging and spectroscopic survey of the sky using a dedicated 2.5 m
telescope \citep{gunn} with $3^{\circ}$ field of view at Apache Point
Observatory in southern New Mexico. The survey is carried out in five
broad filters namely u, g, r, i and z covering the wavelength range
from 3000 to 10000 $A^{\circ}$ \citep{fuku, stout}.

Our present analysis is based on SDSS DR6 \citep{adelman} galaxy
redshift data. The SDSS DR6 includes 9583 $deg^{2}$ imaging and 7425
$deg^{2}$ of spectroscopy. The spectroscopic data includes $1,271,680$
spectra with $790,860$ galaxy redshift. Different selection algorithms
are used for different categories of SDSS targets. We use the Main
Galaxy Sample for the present work. The Main Galaxy Sample target
selection algorithm is detailed in \citealt{strauss}. The Main Galaxy
Sample comprises of galaxies brighter than a limiting r band Petrosian
magnitude $17.77$. We downloaded the data from Catalog Archive Server
(CAS) of SDSS DR6 by Structured Query Language (SQL) search.

For the present analysis we use seven non-overlapping strips each
spanning $90^\circ$ in $\lambda$ and $2^\circ$ in $\eta$, lying
entirely within the survey area of SDSS DR6. $\lambda$ and $\eta$ are
survey co-ordinates described in \citet{stout}. These strips are
identical in sky coverage as the ones used in \citet{pandey1} and are
shown in Figure 1 of that paper. For each strip we have extracted two
different volume limited subsamples in bin1 and bin2 (Table~1 and
Table~2). For each strip the number of galaxies in bin1 and bin2 are
listed in Table~1. The absolute magnitude and redshift limits for bin1
and bin2 are listed in Table~2. 

The analysis using ``Shuffle'' requires us to cut the entire survey
area into squares and shuffle them around. The thickness of the
resulting subsamples increases with redshift. For our analysis we have
considered a smaller region of uniform thickness corresponding to the
value at the lowest redshift. For each luminosity bin the area and
number density are listed in Table~2. For each bin the galaxy number
density varies slightly across the seven strips and the average
density along with the $1-\sigma$ variation is shown in Table~2. For
all the subsamples, the thickness is much smaller than the other two
dimensions and hence it is collapsed along the thickness resulting in
a 2D distribution (Figure \ref{fig:1}).

We study luminosity dependence of statistically significant length
scale of filaments by measuring and comparing it's value in bin1 and
bin2. We label the galaxies in bin1 as Faint and galaxies in bin 2 as
Bright.

We also separately study the colour and morphology dependence in bin 2
(Table~1 and Table~2) by dividing the galaxies in this bin into
Red/Blue and Early/Late type galaxies. The galaxies in bin 1 are not
divided on the basis of their colour or morphology and hence for bin 1
no cut-offs in $u-r$ colour and concentration index $c_i$ are listed
in Table~2.

\begin{table*}{}
\caption{This shows the $(\lambda,\eta)$ range of the seven
 non-overlapping strips. For each strip the table shows the number of
 galaxies in volume limited subsamples bin 1 and bin 2 with absolute
 magnitude and redshift limits given in Table~2.}
\begin{tabular}{|c|c|c|c|c|}
\hline
Strip number & $\lambda(^\circ)$ & $\eta(^\circ) $ & bin 1 &
bin 2 \\
\hline
 1 &$-50\leq\lambda\leq 40$&$9\leq\eta\leq 11$&$845$&$1519$ \\
 2 &$-50\leq\lambda\leq 40$&$11\leq\eta\leq 13$&$743$&$1253$ \\ 
 3 &$-60\leq\lambda\leq 30$ &$13\leq\eta\leq 15$&$624$&$1153$ \\ 
 4 &$-60\leq\lambda\leq 30$&$15\leq\eta\leq 17$&$718$&$1188$ \\
 5 &$-50\leq\lambda\leq 40$&$21.5\leq \eta \leq 23.5$&$774$&$1381$ \\
 6 &$-50\leq\lambda\leq 40$&$24\leq\eta\leq 26$&$799$&$1338$ \\ 
 7 & $-50\leq\lambda\leq 40$&$26\leq\eta\leq 28$&$781$&$1239$ \\ 
\hline
\end{tabular}
\end{table*}

\begin{table*}{}
\caption{This shows the absolute magnitude and redshift limits for the
  different volume limited subsamples analyzed. The area and the
  average galaxy number density with $1-\sigma$ variations from the
  $7$ strips are also shown. The last two column shows the value of
  the galaxy colour $(u-r)_c$ and the value of the concentration index
  $c_{i,c}$ used to divide the data into equal numbers of red/blue and
  elliptical/spiral galaxies respectively. Galaxies in bin 1 are not
  divided and only galaxies in bin2 are divided to obtain different
  samples for red/blue and elliptical/spiral galaxies.
}
\begin{tabular}{|c|c|c|c|c|c|c|}
\hline bin & Absolute Magnitude range & Redshift range & Area\, [
$10^{4}\, h^{-2}\, {\rm Mpc^2}$] & Density \, [$10^{-2}\, h^{2} {\rm
Mpc}^{-2}$] & $(u-r)_c$ & $c_{i,c}$ \\ 
\hline 
bin 1 &$-19 \geq M_r \geq -20$ & $0.028014 \leq z
\leq 0.075635$ & $3.35$ & $2.25 \pm 0.19$ & $-$ & $-$ \\ 
bin 2 &$-20 \geq M_r \geq -21$ &
$0.043657 \leq z \leq 0.114635 $  & $7.47$ &$1.73 \pm 0.15 $ & $2.38$  
& $2.7$\\ \hline
\end{tabular}
\end{table*}

When testing for colour dependence, all the galaxies (Table~1) in the
entire area of the bin 2 (Table~2) are classified as either red or
blue galaxies. The galaxy $u-r$ colours are known to have a bimodal
distribution \citep{strat}. In our analysis we determine a value
$(u-r)_c$ for the colour such that it divides galaxies in bin 2 into
equal number of red (i.e. $u-r>(u-r)_c$) and blue
(i.e. $u-r\leq(u-r)_c$) galaxies. As a consequence the number density
of red and blue galaxies in bin 2 are exactly equal and have a value
half that of the density shown in Table~2.

The morphological classification was carried out using the
concentration index defined as $c_i=r_{90}/r_{50}$ where $r_{90}$ and
$r_{50}$ are the radii containing $90 \%$ and $50 \%$ of the Petrosian
flux respectively. This has been found to be one of the best parameter
to classify galaxy morphology (\citealt{sima}).  Ellipticals are
expected to have a larger concentration indices than spirals. It was
found that $c_i \simeq 3.33$ for a pure de-Vaucouleurs profile
(\citealt{blan2}) while $c_i\simeq 2.3$ for a pure exponential profile
(\citealt{strat}). In bin 2 we have chosen a cutoff $c_{i,c}$
(Table~2) that partitions the galaxies into two equal halves, one
predominantly ellipticals and the other spirals.

\subsection{N-body data}
We simulate the dark matter distribution using a Particle-Mesh (PM)
N-body code. The simulations use $256^3$ particles on a $512^3$ mesh,
and they have a comoving volume $[921.6 h^{-1} {\rm Mpc}]^3$ . We use
$(\Omega_{m0},\Omega_{\Lambda0},h)=(0.27,0.73,0.71)$ for the cosmological
parameters along with a \lcdm power spectrum with spectral index
$n_s=0.96$ and normalization $\sigma_8=0.812$ (\citealt{komat}). 
A ``sharp cutoff'' biasing, scheme \citep{cole} was used to extract
particles from the N-body simulations. Identifying these particles as
galaxies, we have galaxy distributions that are biased relative to the
dark matter. The bias parameter $b$ of each simulated galaxy sample
was estimated using the ratio
\begin{equation}
b=\sqrt{ \frac {\xi_g(r)}{\xi(r)}}
\end{equation}
where $\xi_g(r)$ and $\xi(r)$ are the galaxy and dark matter two-point
correlation functions respectively. This ratio is found to be constant
at length-scales $r \ge 5 h^{-1} {\rm Mpc}$ and we use the average
value over $5-40 h^{-1}{\rm Mpc}$. We use this method to generate
galaxy samples with bias values 1.2, 1.5 and 1.8. We also consider
anti bias and galaxy distributions with bias $b=0.8$ were generated by
adding randomly distributed particles to the dark matter distribution.

\begin{figure}
\rotatebox{0}{\scalebox{.6}{\includegraphics{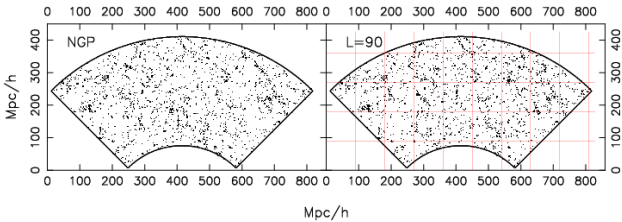}}}
\caption{ This figure exhibits how Shuffle works.  The left panel
  shows the galaxy distribution in a simulated NGP strip from dark
  matter $\Lambda$CDM N-body simulations and the right panel shows a
  shuffled realization generated from the same data using $L=90 h^{-1}
  {\rm Mpc}$. A $90 h^{-1}{\rm Mpc} \, \times \, 90 h^{-1} {\rm Mpc}$
  grid was placed on the mock NGP data, and square blocks lying fully
  inside the survey region were randomly shuffled around to generate
  the data shown in the right panel.  This process destroys all
  coherent structures spanning length-scales larger than $L=90 h^{-1}
  {\rm Mpc}$ in the actual data and filaments larger than this in the
  shuffled data arise from chance alignments. Among the filaments seen
  in the right panel, those which run across the block boundaries have
  formed purely from chance alignments.}
\label{fig:1}
\end{figure}

 The peculiar velocity effects were included to produce galaxy
 distributions in redshift space. We have used three independent
 realisations of the N-body simulations, and for each value of bias we
 have extracted three different mock data-sets from each
 simulation. This gives us a total of nine simulated galaxy
 distribution for each bias values. The simulated galaxy distributions
 (Figure \ref{fig:1}) have the same area, thickness and number
 density as that of the uniform thickness volume limited subsample
 constructed out of equatorial strip ($145^{\circ}<\alpha<236^{\circ},
 -1^{\circ}<\delta<1^{\circ}$) from Northern Galactic Cap (NGP) in
 Paper I. Each strip extends from $235 \, h^{-1} \Mpc$ to $571 \,
 h^{-1} \Mpc$ comoving in the radial direction, has uniform thickness
 of $8.2 \, h^{-1} \, \Mpc$ and contains 1936 galaxies (Table~1 of
 Paper I). We choose to simulate this strip as we have already
 established in Paper I that filaments are statistically significant
 upto length scales $80 \, h^{-1} {\rm Mpc}$ and we want to compare
 this value with that measured from simulations. The simulated data
 were analyzed in exactly the same way as the actual data.

Earlier studies (Paper II) using N-body simulations show that the
filamentarity in the galaxy distribution is sensitive to the area and
galaxy number density of the samples. We want to test whether the
statistically significant length scale of filaments depends on these
factors. We extract regions of identical shape and size similar to the
uniform thickness NGP strip in Paper I from dark matter $\Lambda$CDM
N-body simulations. We label the mock SDSS NGP strip as density 2 and
area 3 in Table~3. Thus effectively the mock SDSS NGP strips, area 3
strips and density 2 strips (Table~3) represent the same galaxy
samples. We then prepare different sets of mock galaxy samples which
have the same area as SDSS NGP strip but different number densities
(density 1, density 3 in Table~3) and other which have the same number
densities as SDSS NGP strip but cover different areas (area 1 and area 2 in
Table~3). For each sample described in Table~3 we extract three
simulated strips from each simulation giving us nine simulated slices.
All these simulated slices are analyzed in exactly the same way as the
actual data.

\begin{table*}{}
\caption{This shows the redshift limits, area and density for the
  different mock subsamples analyzed to test the effect of area and
  density on the statistically significant length scale of filaments.}
\begin{tabular}{|c|c|c|c|}
\hline sample & Redshift range & Area\, [
$10^{4}\, h^{-2}\, {\rm Mpc^2}$] & Density \, [$10^{-2}\, h^{2} {\rm
Mpc}^{-2}$] \\ 
\hline 
area 1 & $0.01\leq z\leq 0.1$ & $6.66$ & $9.09$ \\ 
area 2 & $0.05\leq z\leq 0.15$ & $13.06$ & $9.09$ \\ 
area 3 (NGP strip) & $0.08\leq z\leq 0.2$ & $21.3$ & $9.09$ \\ 
\\ \hline
density 1 & $0.08\leq z\leq 0.2$ & $21.3$ & $4.54$ \\ 
density 2 (NGP strip) & $0.08\leq z\leq 0.2$ & $21.3$ & $9.09$ \\ 
density 3 & $0.08\leq z\leq 0.2$ & $21.3$ & $18.17$ \\
\\ \hline
\end{tabular}
\end{table*}


\begin{figure}
\rotatebox{-90}{\scalebox{.4}{\includegraphics{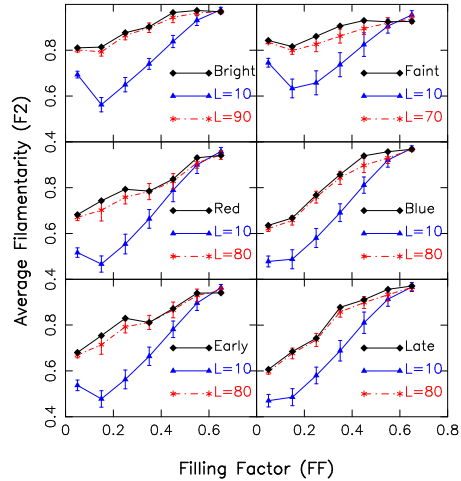}}}
\caption{ This shows the Average Filamentarity as a function of
  Filling Factor ($FF$) for SDSS galaxies in strip 1 (Table~1)
  together with the results for the shuffled data for two values of
  $L$ shown in the figure. Results for different class of galaxies are
  shown in different panels and indicated in each panel. In all the
  panels shuffling with $L=10 h^{-1} \, {\rm Mpc}$ causes a large drop
  in the Average Filamentarity showing the statistical significance of
  the filamentarity at this length-scale.  The data is within the
  $1-\sigma$ error bars of the shuffled realizations for $L=90 h^{-1}
  \, {\rm Mpc}$ (top left panel), $L=70 h^{-1} \, {\rm Mpc}$ (top
  right panel) and $L=80 h^{-1} \, {\rm Mpc}$ (bottom left and right
  panel). The filamentarity is statistically significant up to
  $L_{MAX}= 80, 60$ and $70 h^{-1} \, {\rm Mpc}$ in these cases where
  the actual data lies just above the the $1-\sigma$ error bars and
  are not shown in this figure. For all the larger values of $L$, the
  data remains within the $1-\sigma$ error bars of the shuffled
  realizations indicating that the filaments are not statistically
  significant beyond $L_{MAX}$.  }
\label{fig:2}
\end{figure}

\begin{figure}
\rotatebox{-90}{\scalebox{.4}{\includegraphics{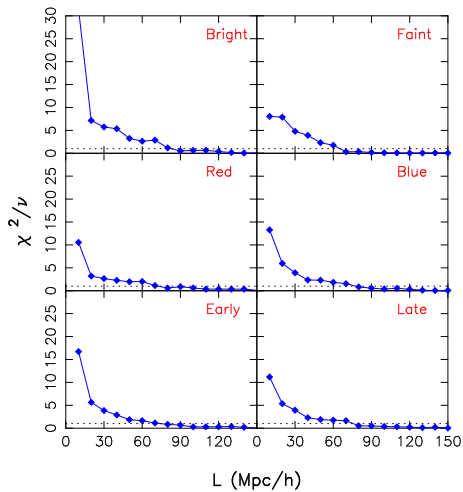}}}
\caption{ This shows $\chi^2/\nu$ at different shuffling lengths for
  different class of SDSS galaxies as indicated in each panel. The
  black dotted line indicates $\chi^2/\nu=1$.}
\label{fig:3}
\end{figure}

\begin{figure}
\rotatebox{-90}{\scalebox{.4}{\includegraphics{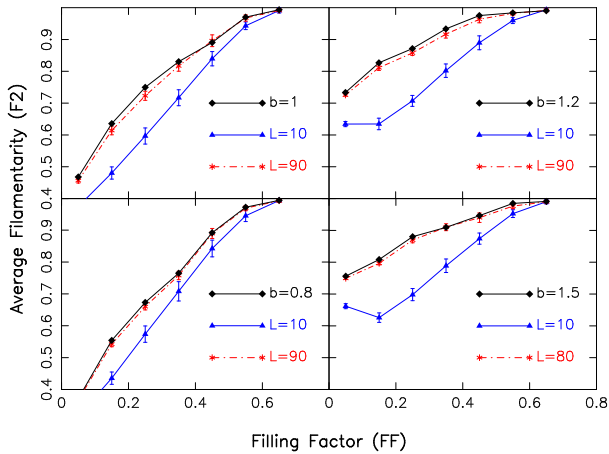}}}
\caption{This shows the results similar to Figure 2 but for a
  simulated SDSS NGP strip with different bias values as indicated in
  each panel.}
\label{fig:4}
\end{figure}

\subsection{Segment Cox Process}
We use the segment Cox process \citep{pons} to test the reliability of
our method in determining the statistically significant length scale
of filaments. This is a controlled point process in which the segments
of length $l$ are scattered with random positions and orientations
within a cube. Points are then randomly distributed on these
segments. The length density of the system of segments is
$L_{v}=\lambda_{s}\, l$, where $\lambda_{s}$ is the mean number of
segments per unit volume. If $\lambda_{l}$ is the mean number of
points on a segment per unit length, then the intensity $\lambda$ of
the resulting point process is, $\lambda=\lambda_{l}\,
L_{v}=\lambda_{l} \, \lambda_{s}\, l$.

For each of the two different segment length ($l=20\, h^{-1} {\rm Mpc}
$ and $l=80\, h^{-1} {\rm Mpc}$) we simulate three realizations of
segment Cox process inside a cube of side length $L=921.6 \, h^{-1}
{\rm Mpc}$. We choose the same box size as the N-body simulations used
in this paper to enable us to extract regions of identical shape and
size similar to the uniform thickness NGP strip used in Paper I. Two
different segment lengths are chosen to represent shorter and longer
filaments. The different parameters used to simulate the segment Cox
process for two different segment lengths are listed in Table~4. We
keep the intensity of the resulting point process same in two cases by
setting $\lambda_{l}$ fixed and adjusting the value of $\lambda_{s}$
with the value of $l$. Thus we have simulated the segment Cox process
for two different values of $l$ and for each values of $l$ the point
process is simulated inside three different boxes. We extract three
different simulated strips from each of the three boxes giving us a
total nine simulated galaxy distribution for each $l$ values. The
simulated strips have the same area, thickness and the number density
as that of the uniform thickness SDSS NGP strip used in Paper I. The
simulated strips are analyzed in exactly the same way as the actual
data.

The simulated 2D strips are constructed out of 3D simulations of
segment Cox process by slicing and projecting and this fact destroys
many segments and shortens many others. We aim to quantify this effect
by generating planar (2D) segment Cox process where all the segments
lie on the same plane. We simulate the 2D segment Cox process over a
planar region which has area and geometry identical to the uniform
thickness SDSS NGP strip used in Paper I. The segments of length $l$
are scattered with random positions and orientations within this
planar region. Points are then randomly distributed on these
segments. The intensity $\lambda$ of the resulting point process is,
$\lambda=\lambda_{l} \, \lambda_{s}\, l$ where $\lambda_{s}$ is the
mean number of segments per unit area, $\lambda_{l}$ is the mean
number of points on a segment per unit length and $l$ is the length of
the segments. We simulate 9 such realizations of 2D segment Cox
process for each of the two segment lengths ($l=20\, h^{-1} {\rm Mpc}
$ and $l=80\, h^{-1} {\rm Mpc}$). The different parameters used to
simulate the 2D segment Cox process for two different segment lengths
are listed in Table~5. We now randomly extract exactly same number of
points as there are in the uniform thickness SDSS NGP strip used in
Paper I from each of the 9 realizations of 2D segment Cox process.
This gives us a total nine simulated galaxy distribution for each
segment lengths (Table~5). The simulated strips have same area,
geometry and number density as the SDSS NGP strip. We analyze these
simulated strips exactly the same way as the actual data.

\begin{table*}{}
\caption{This shows different parameters used to simulate 3D segment
  Cox process.  }
\begin{tabular}{|c|c|c|c|}
\hline $l$&$\lambda_s$&$\lambda_l$&$\lambda$\\ (Segment length)&(Mean no. of
segments per unit volume)&(Mean no. of points on a segment per unit
length)&(Intensity)\\\, [$h^{-1}\, {\rm Mpc}$]&\, [$h^{3}\, {\rm Mpc^{-3}}$]&\, [$h \, {\rm Mpc^{-1}}$]&\, [$h^{3}\, {\rm Mpc^{-3}}$]\\

\hline 
$20$ & $0.001278$ & $0.8$ & $0.020441$ \\ 
$80$ & $0.000319$ & $0.8$ & $0.020441$ \\ \hline
\end{tabular}
\end{table*}

\begin{table*}{}
\caption{This shows different parameters used to simulate 2D segment
  Cox process.  }
\begin{tabular}{|c|c|c|c|}
\hline $l$&$\lambda_s$&$\lambda_l$&$\lambda$\\(Segment length)&(Mean no. of
segments per unit area)&(Mean no. of points on a segment per unit
length)&(Intensity)\\ \, [$h^{-1}\, {\rm Mpc}$]&\, [$h^{2}\, {\rm Mpc^{-2}}$]&\, [$h \, {\rm Mpc^{-1}}$]&\, [$h^{2}\, {\rm Mpc^{-2}}$]\\
\hline 
$20$ & $0.037612$ & $0.8$ & $0.601791$ \\ 
$80$ & $0.009403$ & $0.8$ & $0.601791$ \\ \hline
\end{tabular}
\end{table*}

\subsection{Method of Analysis}

All the strips that we have analyzed are nearly two dimensional.  The
strips were all collapsed along the thickness (the smallest dimension)
to produce 2D galaxy distributions. We use the 2D ``Shapefinder''
statistic \citep{bharad1} to quantify the average filamentarity of the
patterns in the resulting galaxy distribution.  A detailed discussion
is presented in Paper II , and we present only the salient features
here. The reader is referred to \citet{sahni} for a discussion of
Shapefinders in three dimensions.

 The galaxy distribution is represented as a set of 1s on a 2-D
 rectangular grid of spacing $1 \, h^{-1} {\rm Mpc} \times 1 \,h^{-1}
 {\rm Mpc}$, empty cells are assigned a value $0$ and filled cells are
 assigned a value $1$. We identify connected filled cells using the
 `Friends-of-Friend' (FOF) algorithm. The filamentarity of each
 cluster is quantified using the Shapefinder ${\cal F}$ defined as
\begin{equation}
{\cal F} = \frac{(P^2 - 16 S)}{(P-4 l)^2}
\end{equation}
 where $P$ and $S$ are respectively the perimeter and the area of the
 cluster, and $l$ is the grid spacing. The  Shapefinder ${\cal F}$
 has  values $0$ and $1$ for a square and filament respectively, and
 it assumes intermediate values as  a  square is deformed to a
 filament. We use the average filamentarity  
\begin{equation} 
F_2 = {\sum_{i} {\cal S}_i^2 {\cal F}_i\over\sum_{i}{\cal S}_i^2} \,. 
\end{equation}
to asses the overall filamentarity of  the clusters in the galaxy
distribution. 

The distribution of 1s corresponding to the galaxies is sparse. Only
$\sim 1 \%$ of the cells contain galaxies and there are very few
filled cells which are interconnected.  As a consequence FOF fails to
identify the large coherent structures which corresponds to filaments
in the galaxy distribution. We overcome this by successively
coarse-graining the galaxy distribution. In each iteration of
coarse-graining all the empty cells adjacent to a filled cell
(i.e. cells at the 4 sides and 4 corners of a filled cell) are
assigned a value $1$ . This causes clusters to grow, first because of
the growth of individual filled cells, and then by the merger of
adjacent clusters as they overlap. Coherent structures extending
across progressively larger length-scales are identified in
consecutive iterations of coarse-graining.  So as not to restrict our
analysis to an arbitrarily chosen level of coarse-graining, we study
the average filamentarity after each iteration of coarse-graining. The
filling factor $FF$ quantifies the fraction of cells that are filled
and its value increases from $\sim 0.01$ and approaches $1$ as the
coarse-graining proceeds. We study the average filamentarity $F_2$ as
a function of the filling factor $FF$ (Figure \ref{fig:2}) as a
quantitative measure of the filamentarity at different levels of
coarse-graining. The values of $FF$ corresponding to a particular
level of coarse-graining shows a slight variation from strip to strip.
In order to combine and compare the results from different strips, for
each strip we have interpolated $F_2$ to $7$ values of $FF$ at an
uniform spacing of $0.1$ over the interval $0.05$ to
$0.65$. Coarse-graining beyond $FF \sim 0.65$ washes away the
filaments and hence we do not include this range for our analysis.

We use a statistical technique ``Shuffle'' to determine the largest
length-scale at which the filamentarity is statistically significant.
A grid with squares blocks of side $L$ is superposed on the original
data slice (Figure \ref{fig:1}). Blocks of data which lie entirely
within the slice are then randomly interchanged, with rotation,
repeatedly, to form a new shuffled slice. The shuffling process
eliminates coherent features in the original data on scales larger
than $L$, keeping clustering at scales below $L$ nearly identical to
the original data.  All the structures spanning length-scales greater
than $L$ that exist in the shuffled slices are the result of chance
alignments. At a fixed value of $L$, the average filamentarity in the
original sample will be larger than in the shuffled data only if the
actual data has more filaments spanning length-scales larger than $L$,
than that expected from chance alignments. The largest value of $L$,
$L_{\rm MAX}$, for which the average filamentarity of the shuffled
slices is less than the average filamentarity of the actual data gives
us the largest length-scale at which the filamentarity is
statistically significant. Filaments spanning length-scales larger
than $L_{\rm MAX}$ arise purely from chance alignments.

 For each value of $L$ we generated $24$ different realization of the
 shuffled slices.  To ensure that the edges of the blocks which are
 shuffled around do not cut the actual filamentary pattern at exactly
 the same place in all the realizations of the shuffled data, we
 randomly shifted the origin of the grid used to define the blocks.
 The values of FF and $F_2$ in the 24 realizations differ from one
 another and from the actual data at the same stage of
 coarse-graining. So as to be able to quantitatively compare the
 shuffled realizations with the actual data, we interpolate the values
 of $F_2$ in the shuffled realization at the same values of FF as
 interpolated for the actual data.  The mean $\bar{F_2}[{\rm
     Shuffled}]$ and the variance $(\Delta F_2[{\rm Shuffled}])^2$ of
 the average filamentarity was determined for the shuffled data at
 each value of FF using the 24 realizations.  The difference between
 the filamentarity of the shuffled data and the actual data was
 quantified using the reduced $\chi^2$ per degree of freedom
\begin{equation}
\frac{\chi^2}{\nu}=\frac{1}{N_p} \sum_{a=1}^{N_p}
 \frac{(F_2[{\rm Actual}]-\bar{F_2}[{\rm Shuffled}]))_a^2 }
{(\Delta  F_2[{\rm Shuffled}])_a^2} 
\end{equation}
where the sum is over different values of the filling factor FF. 

\section{Results and Conclusions}

We use ``Shuffle'' to determine the statistically significant length
scale of filaments in the seven SDSS strips in magnitude bin 1 and bin
2 (Table~1 and Table~2). We label bin 1 as Faint and bin 2 as
Bright. The top two panels of Figure \ref{fig:2} shows the results of
``Shuffle'' on strip 1 from bin 1 and bin 2 respectively. In top two
panels of Figure \ref{fig:2} we see that shuffling the data with
$L=10$ causes a large drop in the average filamentarity ($F_2$). The
average filamentarity increases as the value of $L$ increases and the
value of average filamentarity of the unshuffled slice comes within $1
-\sigma$ error bars when the data is shuffled with $L=90$ and $L=70$ for
Bright and Faint galaxies respectively. Difference between the average
filamentarity of shuffled and unshuffled slice is quantified by
$\chi^2/\nu$ and the top two panels of Figure \ref{fig:3} shows
$\chi^2/\nu$ as a function of $L$. In top two panels of Figure
\ref{fig:3} we see that $\chi^2/\nu$ approaches 1 at $L=90$ for Bright
and $L=70$ for Faint galaxies. So in strip 1 from bin1 and bin 2 the
filaments are statistically significant upto length scales $80 \,
h^{-1} {\rm Mpc}$ and $60 \, h^{-1} {\rm Mpc}$ respectively. The
results from the other strips are not shown here. Averaging the
results from all the nine strips we find that the filaments are
statistically significant upto length scales $84 \pm 16 \, h^{-1}
{\rm Mpc}$ and $71 \pm 16 \, h^{-1} {\rm Mpc}$ in bin 1 and bin
2 respectively.

The right and left middle panels of Figure \ref{fig:2} shows the
results of ``Shuffle'' for Red and Blue galaxies (Table~2) from strip
1 in bin 2. The right and left middle panels of Figure \ref{fig:2}
shows that a large drop in the average filamentarity is observed when
the data is shuffled with $L=10$. In both the panels we see that
shuffling the data with $L=80$ increase the average filamentarity and
brings it back in agreement with the actual data. In two middle panels
of Figure \ref{fig:3} we see that in both class of galaxies
$\chi^2/\nu$ approaches 1 at $L=80$ establishing that for both Red and
Blue galaxies the filaments are statistically significant upto length
scales $70 \, h^{-1} {\rm Mpc}$. Here we have not shown the results from
the other strips. Combining results from all the nine strips we
find that the filaments are statistically significant upto length
scales $77 \pm 10 \, h^{-1} {\rm Mpc}$ and $71 \pm 10 \,
h^{-1} {\rm Mpc}$ for Red and Blue galaxies respectively.

We show the results of ``Shuffle'' for the Early and Late type
galaxies (Table~2) in the right and left bottom panels of Figure
\ref{fig:2}. Here also we see that the average filamentarity decreases
when the data is shuffled with $L=10$ and it increases and matches
with the actual data when the data is shuffled with $L=80$. The
$\chi^2/\nu$ approaches 1 at $L=80$ for both Early and Late type
galaxies as shown in two bottom panels of Figure \ref{fig:3}. This
establishes that for both Early and Late type galaxies the filaments
are statistically significant upto length scales $70 \, h^{-1} {\rm
  Mpc}$ and not beyond. We have shown here only the results for strip
1 in bin 2. Combining the results from all the strips the filaments
are found to be statistically significant upto length scales $75 
\pm 10  \, h^{-1} {\rm Mpc}$ and $74 \pm 17  \, h^{-1} {\rm
  Mpc}$ for Early and Late type galaxies respectively.

It is to be noted in middle and bottom two panels of Figure
\ref{fig:2} that the Red and Early type galaxies have higher degree of
filamentarity as compared to Blue and Late type galaxies at smaller
values of filling factor. We reported this effect in
\citet{pandey1}. It is interesting to note that although the average
filamentarity ($F_2$) depends on luminosity, colour and morphology of
galaxies, the statistically significant length scale of filaments does
not depend on these galaxy properties.

\begin{figure}
\rotatebox{-90}{\scalebox{.4}{\includegraphics{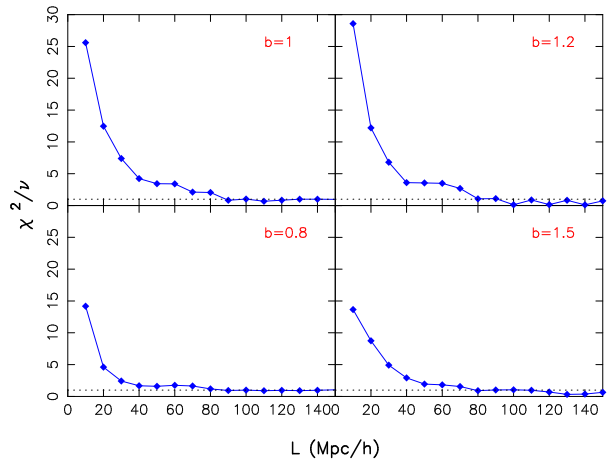}}}
\caption{This shows $\chi^2/\nu$ at different shuffling lengths for
  a simulated SDSS NGP strip with different bias values as indicated in
  each panel. The black dotted line indicates $\chi^2/\nu=1$.}
\label{fig:5}
\end{figure}

In Figure \ref{fig:4} we show the results of ``Shuffle'' in simulated
SDSS NGP strips with different biases. The top left panel show the
result for the dark matter distribution from $\Lambda$CDM N-body
simulations. In this case galaxies are assumed to exactly trace the
dark matter and the bias parameter $b=1$. We see that shuffling the
simulated data with $L=10$ causes a large drop in the average
filamentarity of the simulated galaxy distribution. With increasing L
the value of average filamentarity of the shuffled slice slowly
approaches the values corresponding to the unshuflled data. It is
found that at $L=90$ the two are within $1 -\sigma$ error bars. The top
right panel shows the result for b=1.2. We see a very similar result
and the unshuflled and shuffled data agrees well when the data is
shuffled with $L=90$. In top left and right panels of Figure
\ref{fig:5} we see that $\chi^2/\nu$ approaches 1 at $L=90$ for both
$b=1$ and $b=1.2$ and hence filaments are statistically significant
upto length scales $80 \, h^{-1} {\rm Mpc}$ for both cases. The
results are shown for one single mock SDSS strip. Averaging the
results from all the nine strips it is found that for b=1 and b=1.2
filaments are statistically significant upto length scales $93  \pm
16  \, h^{-1} {\rm Mpc}$ and $92  \pm 15  \, h^{-1} {\rm Mpc}$
respectively.  The bottom left panel of Figure \ref{fig:4} shows the
result of shuffling on simulated SDSS strips for a bias b=0.8. As
usual shuffling the data with $L=10$ causes a drop in average
filamentarity of the shuffled slice but the drop is smaller as
compared to other bias values. As L increases the average
filamentarity slowly grows and approaches the values corresponding to
the unshuffled data at $L=90$. In the bottom right panel of Figure
\ref{fig:4} results are shown for a high bias b=1.5. Shuffling shows
similar results for b=1.5 but a relatively large drop is seen when the
data is shuffled with $L=10$. With increasing L the average
filamentarity grows relatively faster as compared to b=0.8 and finally
the average filamentarity in the shuffled data levels up with the
unshuffled data when $L=80$ is used for shuffling. In the bottom left
and right panels of Figure \ref{fig:5} it is found that $\chi^2/\nu$
approaches 1 at $L=90$ and $L=80$ indicating that filaments are
statistically significant upto length scales $80 \, h^{-1} {\rm Mpc}$
and $70 \, h^{-1} {\rm Mpc}$ for b=0.8 and b=1.5 respectively. These
are the results shown for a single mock SDSS strip and by combining
the results from the other strips we get that the filaments are
statistically significant upto length scales $91 \pm 13  \,
h^{-1} {\rm Mpc}$ and $90 \pm 15 \, h^{-1} {\rm Mpc}$ for $b=0.8$
and $b=1.5$ respectively.

\begin{figure}
\rotatebox{-90}{\scalebox{.4}{\includegraphics{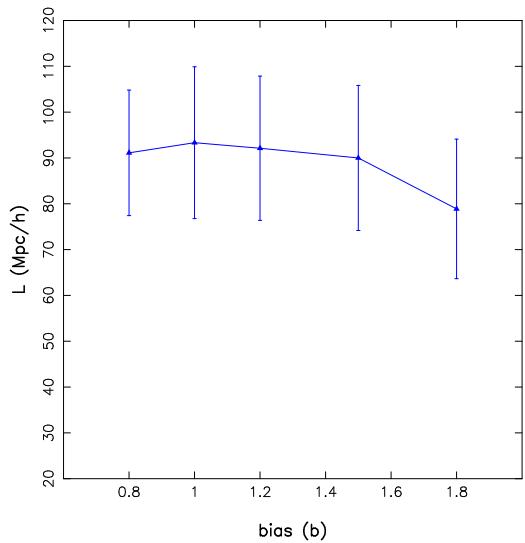}}}
\caption{This shows the statistically significant length scale of
  filaments as a function of bias b. The error bars are $1-\sigma$
  error bars measured from nine mock galaxy catalogues.}
\label{fig:6}
\end{figure}

\begin{figure}
\rotatebox{-90}{\scalebox{.4}{\includegraphics{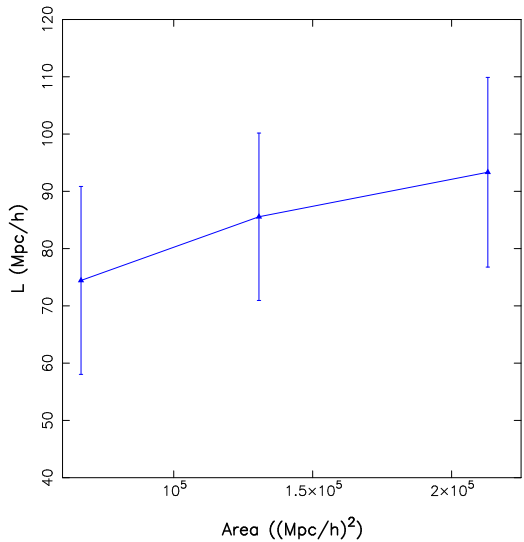}}}
\caption{This shows the statistically significant length scale of
  filaments as a function area of the samples described in Table~3. The
  error bars are $1 - \sigma$ error bars measured from nine mock galaxy
  catalogues. }
\label{fig:7}
\end{figure}

\begin{figure}
\rotatebox{-90}{\scalebox{.4}{\includegraphics{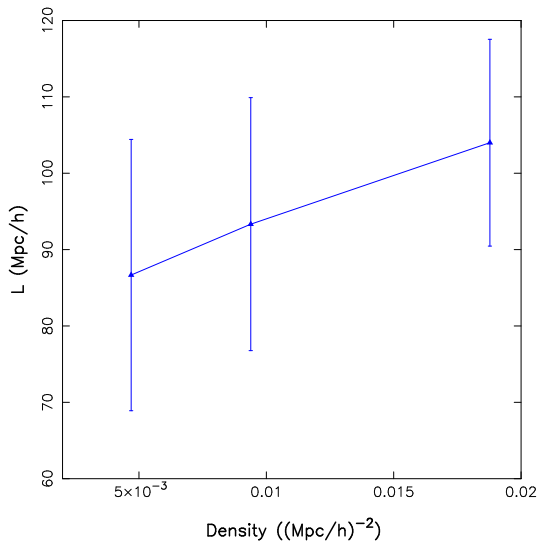}}}
\caption{This shows the statistically significant length scale of
  filaments as a function galaxy number density of the samples
  described in Table~3. The error bars are $1 - \sigma$ error bars
  measured from nine mock galaxy catalogues. }
\label{fig:8}
\end{figure}

It is to be noted in Figure \ref{fig:4} that as the bias $b$ is
increased the average filamentarity ($F_2$) increases at smaller $FF$.
We reported this effect in \citet{bharad3}. But interestingly the
shuffling length at which the avergae filamentarity of the unshuflled
and shuffled data agrees does not depend on bias. Figure \ref{fig:6}
shows the statistically significant length scale of filaments as a
function of bias. This Figure shows that the statistically significant
length scale of filaments is nearly independent of bias. The mean
value is $~90 \, h^{-1} {\rm Mpc} $ for the bias values
$b=0.8-1.5$. It tends to decrease for a high value of bias
$b=1.8$. This could be possibly related to the fact that at very high
values of bias we preferentially identify only very high density
regions which are mostly related to cluster like structures rather
than filaments.

Earlier in Paper I we established that the filaments are statistically
significant upto length scales of $80 \, h^{-1} {\rm Mpc}$. This
measured value lies well within $1-\sigma$ error bars of the measured
values for mock galaxy distribution from $\Lambda$CDM N-body
simulations. We conclude that the $\Lambda$CDM is consistent with SDSS
observations.

The galaxy samples constructed over different magnitude ranges have
different area and galaxy number density (Table~2). Earlier analysis
in Paper II shows that the average filamentarity of the galaxy
distribution depends on the area and galaxy number density of the
galaxy samples and filamentarity between different galaxy samples can
only be compared when the samples have same area, geometry and number
density of galaxies. So it is important to know if the statistically
significant length scale of filaments depends on these factors before
we can compare it's measured values among various galaxy samples
having different areas and galaxy number densities.

We prepare two different set of mock galaxy samples namely area 1 and
area 2 (Table~3) which have the same galaxy number density as the SDSS
NGP strip (area 3) but cover different redshift ranges and hence have
different areas as listed in Table~3. We extract three mock galaxy
samples from each of the $\Lambda$CDM dark matter simulation for each
set of galaxy samples area 1 and area 2. We determine the
statistically significant length scale of filaments for these galaxy
samples. The analysis for the mock SDSS NGP strip which we name as
area 3 in Table~3 are already done. We plot the statistically
significant length scale of filaments as a function of the area of the
galaxy samples in Figure \ref{fig:7}. From this figure we see that the
samples of smaller area tend to have a smaller length scale of
filaments. But this trend is weak and the size of the error bars are
quite large. We conclude that the statistically significant length
scale of filaments are nearly independent of the area of the samples.

We next prepare two different set of mock galaxy samples namely
density 1 and density 3 (Table~3) which have the same area as the SDSS
NGP strip (density 2) but have different number densities listed in
Table~3. We extract three mock galaxy samples from each of the
$\Lambda$CDM dark matter simulation for each set of galaxy samples
density 1 and density 3. We determine the statistically significant
length scale of filaments for these galaxy samples. The analysis for
the mock SDSS NGP strip which we name as density 2 in Table~3 are
already done. We plot the statistically significant length scale of
filaments as a function of the number density of galaxies of the
samples in Figure \ref{fig:8}. We see a trend of low density samples
to have a smaller length scale of filaments. But the error bars are
quite large and the dependence is weak. We conclude that the
statistically significant length scale of filaments weakly depends on
the number density of galaxies. In a recent work with SDSS LRG samples
\citep{pandey4} having much larger area and lower density we find that
the statistically significant length scale of filaments does not
change much as compared to it's value measured in SDSS Main galaxy
samples. Thus the statistically significant length scale of filaments
is a more robust statistics than the average filamentarity. It does
not depend on properties (size and number density) of galaxy samples ,
and physical properties of galaxies. It emerges as a robust measure of
the large scale structures. The statistically significant length scale
of filaments could be also thought of as an indicator of the scale
beyond which the galaxy distribution become homogeneous because it
tells us the length scale beyond which no coherent structures exist in
the galaxy distribution. Multifractal analysis in SDSS DR1
\citep{yadav} and recently in SDSS DR6 \citep{sarkar} show that the
galaxy distribution becomes homogeneous at a length scale between $60$
to $70 h^{-1} \, {\rm Mpc}$. \citet{yadav} considered the transition
to homogeneity in $\Lambda$CDM simulations with different bias
$b=1,1.6,2$ and find that irrespective of the bias values the
distribution become homogeneous at a length scale between $60$ to $70
h^{-1} \, {\rm Mpc}$. These results are consistent with our findings.

Finally we test the reliability of our method by applying it to
different controlled samples (Table~4) of 3D segment Cox process. We
use segments of two different length $l=20\, h^{-1} {\rm Mpc}$ and
$l=80\, h^{-1} {\rm Mpc}$ for simulating the 3D segment Cox process.
Two different sets of 2D mock samples drawn from these 3D simulations
(Table~4) are separately analyzed. The results are shown in top two
panels of Figure \ref{fig:9} and Figure \ref{fig:10}. We see in top
left panel of Figure \ref{fig:9} that filamentarity decrease by a very
small amount when the mock samples having $l=20\, h^{-1} {\rm Mpc}$
are shuffled with $L=10$ and there are virtually no drop in
filamentarity when the data is shuffled with $L=20$. The top right
panel of the same figure shows the result of ``Shuffle'' on cotrolled
samples of segment Cox process with segment length $l=80\, h^{-1} {\rm
  Mpc}$. We see a significant drop in the average filamentarity when
the data is shuffled with $L=10$ and the average filamentarity finally
saturates to the values corresponding to the unshuffled data when
$L=60$ is used for shuffling. The $\chi^2/\nu$ as a function of $L$
for these two cases are shown in top two panels of Figure
\ref{fig:10}. We see that $\chi^2/\nu$ approaches 1 at $L=20$ and
$L=60$ in these two cases respectively. This indicates that for these 2D
mock samples having $l=20\, h^{-1} {\rm Mpc}$ and $l=80\, h^{-1} {\rm
  Mpc}$, the filaments are found to be statistically significant upto
length scales $10 \, h^{-1} {\rm Mpc}$ and $50 \, h^{-1} {\rm Mpc}$
respectively. Here the results are shown only for a single simulated
mock sample. By averaging the results from all the nine simulated mock
samples we find that in the above two cases the filaments are
statistically significant upto length scales $18 \pm 7 \, h^{-1} {\rm
  Mpc}$ and $52 \pm 11 \, h^{-1} {\rm Mpc}$ respectively. We note that
the values of statistically significant length scale of filaments
recovered by our method for straight segments simulated here are a
little shorter but close to the original values of $l$ used in the
simulations. The simulations are carried out in 3D boxes and our mock
samples are the 2D projections of nearly two dimensional samples drawn
from these boxes. This possibly destroys and shortens a lot of
segments and the effects are expected to be more prominent for longer
segments as reflected in our results. This effect indicates that also
for the 2D real galaxy samples from SDSS which are constructed out of
a 3D galaxy distribution, Shuffle possibly underestimates the values
of statistically significant length scale of the real 3D filaments.

\begin{figure}
\rotatebox{0}{\scalebox{.6}{\includegraphics{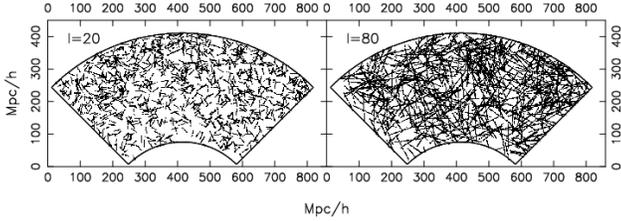}}}
\caption{ This shows a simulation of 2D segment Cox process with two
  different segment length $l=20\, h^{-1} {\rm Mpc}$ and $l=80\,
  h^{-1} {\rm Mpc}$ (as indicated in each panel) over a region which
  has identical geometry to the SDSS NGP equatorial strip. These
  simulated strips have same values of $l$ and $\lambda_{l}$ listed in
  Table~5 but different values of $\lambda_{s}$ are chosen so as
  to lower the density of the segments and make them visibly
  distinguishable in this plot.}
\label{fig:11}
\end{figure}

\begin{figure}
\rotatebox{-90}{\scalebox{.4}{\includegraphics{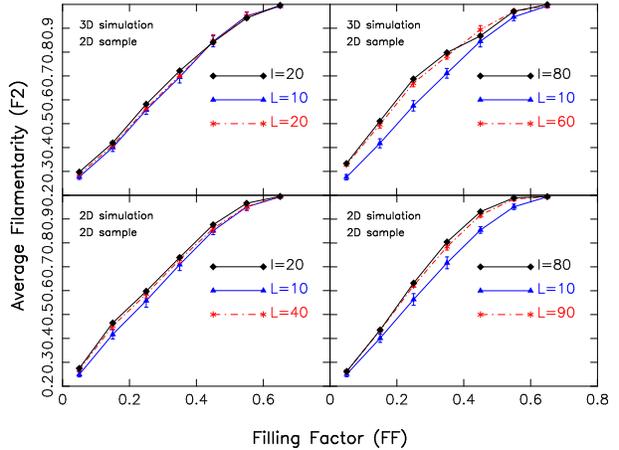}}}
\caption{ This shows the Average Filamentarity as a function of
  Filling Factor ($FF$) for a mock SDSS NGP strip constructed out of
  3D simulation of segment Cox process (Table~4) and 2D simulation
  of segment Cox process (Table~5) together with the results for the
  shuffled data for two values of $L$ shown in the figure. The 2D
  simulations of segment Cox process contain all the segments in the
  same plane and there are no effects because of the slicing process
  or the projections which arises when samples are constructed out of
  3D simulations of segment Cox process. We see in this plot that
  ``Shuffle" is able to recover the length of the longer segments more
  efficiently when samples are constructed out of 2D simulations
  instead of 3D simulations of segment Cox process. }
\label{fig:9}
\end{figure}

\begin{figure}
\rotatebox{-90}{\scalebox{.4}{\includegraphics{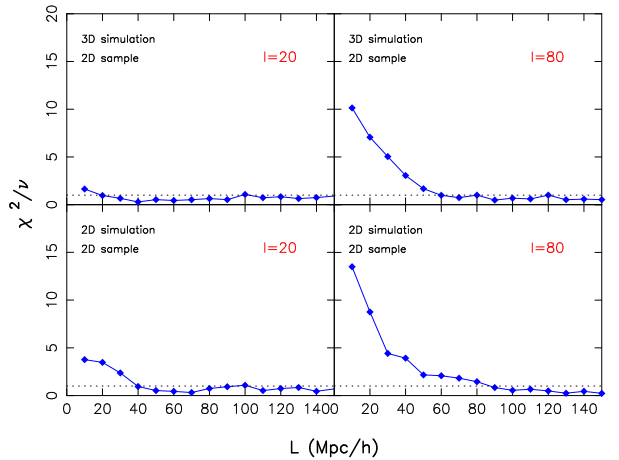}}}
\caption{ This shows $\chi^2/\nu$ at different shuffling lengths ($L$)
  for a mock SDSS NGP strip constructed out of 3D segment Cox process
  and 2D segment Cox process simulated with two different segment
  length $l=20\, h^{-1} {\rm Mpc}$ and $l=80\, h^{-1} {\rm Mpc}$. The
  black dotted line indicates $\chi^2/\nu=1$.}
\label{fig:10}
\end{figure}

The effects of slicing and projections can be avoided if we simulate
the segment Cox process in 2D as in this case the segments lie on the
same plane (Figure \ref{fig:11}). We apply Shuffle to a set of mock 2D
strips constructed out of simulations of 2D segment Cox process
(Table~5). We use two different segment lengths $l=20\, h^{-1} {\rm
  Mpc}$ and $l=80\, h^{-1} {\rm Mpc}$ to simulate the 2D segment Cox
process. We use the same segment lengths as used in the simulations of
3D segment Cox process so as to test the effect of slicing and
projection. We show the results in bottom two panels of Figure
\ref{fig:9} and Figure \ref{fig:10}. Bottom left panel of Figure
\ref{fig:9} shows that when the mock samples with $l=20\, h^{-1} {\rm
  Mpc}$ are shuffled with $L=10$ the average filamentarity of the
shuffled data falls below that of the unshuffled data. The average
filamentarity of the shuffled data saturates to the unshuffled data
when the data is shuffled with $L=40$. The bottom right panel of
Figure \ref{fig:9} shows a significant drop in the average
filamentarity when the mock samples with $l=80\, h^{-1} {\rm Mpc}$ are
shuffled with $L=10$. The average filamentarity of the shuffled data
increases slowly with increasing shuffling length and finally
saturates to the unshuffled data at shuffling length $L=90$. We see in
bottom two panels of Figure \ref{fig:10} the $\chi^2/\nu$ as a
function of $L$ approaches 1 at $L=40$ and $L=90$ in these two cases
respectively. This indicates that for these controlled samples of 2D
segment Cox process having $l=20\, h^{-1} {\rm Mpc}$ and $l=80\,
h^{-1} {\rm Mpc}$, the filaments are statistically significant upto
length scales $30 \, h^{-1} {\rm Mpc}$ and $80 \, h^{-1} {\rm Mpc}$
respectively. We have shown the results for a single simulated strip
in these plots. We combine the results from all the nine simulated
strips and find that the filaments are statistically significant upto
length scales $25 \pm 8 \, h^{-1} {\rm Mpc}$ and $82 \pm 24 \, h^{-1}
{\rm Mpc}$ for simulations of 2D segment Cox process with segment
length $l=20\, h^{-1} {\rm Mpc}$ and $l=80\, h^{-1} {\rm Mpc}$
respectively.  We note that our method can successfully recover better
the length of inputted segments when the mock samples are drawn from
simulations of 2D segment Cox process where the effects of slicing and
projections are absent. There would be another effect due to the
boundaries of the samples which again shortens the length of some
segments near the boundaries but this effect is not so important in
our analysis as the extent of the samples are much larger than the
length of the inputted segments. This analysis suggets that the actual
length scale of the real 3D filaments can be recovered with Shuffle if
the analysis is extended in 3D. We propose to take up this in a future
work.

In conclusion we note that although the average filamentarity of the
galaxy distribution depends on different galaxy properties, the
statistically significant length scale of filaments does not depend on
luminosity, colour and morphology of galaxies. We find that the
measured values of the statistically significant length scale of
filaments in SDSS are consistent with that measured from $\Lambda$CDM
simulations. We considered the $\Lambda$CDM model with different
values of bias and find that the statistically significant length
scale of filaments is nearly independent of bias. Different class of
galaxies are differently biased with respect to underlying dark matter
distribution and this result is possibly related to the fact that
statistically significant length scale of filaments is nearly the same
for different class of galaxies. Unlike the average filamentarity, the
statistically significant length scale of filaments is nearly
independent of the area and galaxy number density of the samples. This
establishes the statistically significant length scale of filaments as
a robust statistics of galaxy distribution.

\section{Acknowledgment}
BP acknowledges Somnath Bharadwaj for useful discussions and
suggestions. Thanks are due to an anonymous referee, whose comments
and suggestions led to significant improvement of the paper.  The SDSS
DR6 data was downloaded from the SDSS skyserver
http://cas.sdss.org/dr6/en/ .

    Funding for the creation and distribution of the SDSS Archive has been 
provided by the Alfred P. Sloan Foundation, the Participating 
Institutions, the National Aeronautics and Space Administration, the 
National Science Foundation, the U.S. Department of Energy, the Japanese 
Monbukagakusho, and the Max Planck Society. The SDSS Web site is 
http://www.sdss.org/.

    The SDSS is managed by the Astrophysical Research Consortium (ARC) for 
the Participating Institutions. The Participating Institutions are The 
University of Chicago, Fermilab, the Institute for Advanced Study, the 
Japan Participation Group, The Johns Hopkins University, the Korean 
Scientist Group, Los Alamos National Laboratory, the Max-Planck-Institute 
for Astronomy (MPIA), the Max-Planck-Institute for Astrophysics (MPA), New 
Mexico State University, University of Pittsburgh, Princeton University, 
the United States Naval Observatory, and the University of Washington.

\end{document}